 \documentclass[12pt]{iopart}

\usepackage{graphicx}
\usepackage{dcolumn}
\usepackage{bm}
\usepackage{epstopdf}

 \newcommand{\hcc}{{\rm hcc}}
  
  \newcommand{\vsun}{{v_c^\ast}}

\newcommand{\SA}{{\rm SA}}

\newcommand{\bA}{ A}
\newcommand{\bC}{C}

\newcommand{\dtwee}{{\rm d}^2}
\newcommand{\ddrie}{{\rm d}^3}
\renewcommand{\dtwee}{{\rm d}}
\renewcommand{\ddrie}{{\rm d}}

\newcommand{\myskip}[1]{}

\newcommand{\eps}{\varepsilon}
\renewcommand{\d}{{\rm d}}

\newcommand{\km}{{\rm km}}
\renewcommand{\H}{{\rm H}}
\newcommand{\He}{{\rm He}}

\newcommand{\SagA}{SagA$^\ast$}

\newcommand{\SMC}{{\rm SMC}}
\newcommand{\DSA}{d_{\rm SA}}
\newcommand{\gas}{{\rm gas}}
\newcommand{ \pix}{{\rm pix}}

\newcommand{\barR}{{\bar R}}
\newcommand{\barM}{{\bar M}}

\newcommand{\BEQ}{\begin{eqnarray}}
\newcommand{\EEQ}{\end{eqnarray}}
\newcommand{\BEA}{\begin{eqnarray}}
\newcommand{\EEA}{\end{eqnarray}}

\newcommand{\m}{{\rm m}}

\newcommand{\s}{{\rm s}}
\newcommand{\p}{\partial}
\newcommand{\K}{{\rm K}}
\newcommand{\pc}{{\rm pc}}
\newcommand{\jc}{{\rm jc}}
\newcommand{\bd}{{\mu{\rm bd}}}

\newcommand{\LCDM}{{$\Lambda${\rm CDM}}}
\newcommand{\mBD}{$\mu${\rm BD}}
\newcommand{\mBDs}{$\mu${\rm BDs}}

\begin{document}
\title{Do the Herschel cold clouds in the Galactic halo embody its dark matter?}

\author{Theo M. Nieuwenhuizen, Erik F. G. van Heusden and Matthew T. P. Liska}
\address{Institute for Theoretical Physics, University of Amsterdam,
Science Park 904, \\ P.O. Box 94485,  1090 GL  Amsterdam, The Netherlands }

\hspace{1.6cm} email: {t.m.nieuwenhuizen@uva.nl}

\begin{abstract}
Recent Herschel/SPIRE maps of the Small and Large Magellanic Clouds  (SMC, LMC) exhibit in each thousands of clouds.
Observed at 250 microns, they must be cold, $T\sim15$ K, hence the name ``Herschel cold clouds'' (HCCs).
From the observed rotational velocity profile of the Galaxy and the assumption of spherical symmetry, its
mass density is modeled in a form close to that of an isothermal sphere. 
If the HCCs constitute a certain fraction of it,  their angular size distribution has a specified shape.
A fit to the data deduced from the SMC/LMC maps supports this and yields 1.7 pc for  their average radius.
There are so many HCCs that they will make up all the missing Halo mass density if there is spherical symmetry and their average mass is 
of order $10,000M_\odot$. This compares with the Jeans mass of circa  $40,000M_\odot$ and puts forward that the HCCs are in fact Jeans clusters, 
constituting  all the Galactic dark matter and much of its missing baryons,
a conclusion deduced before from a different field of the sky (Nieuwenhuizen, Schild and Gibson 2011).
A preliminary analysis of the intensities yields that the Jeans clusters themselves may consist of some billion MACHOs of a few dozen Earth masses. 
With a size of dozens of solar radii, they would mostly not lens but cause occultation of stars in the LMC, SMC and towards the Galactic center, 
and may thus have been overlooked in microlensing.
\end{abstract}

\hspace{1.6cm} PACS numbers:\\ \indent
\hspace{1.6cm} 95.35.+d	Dark matter (stellar, interstellar, galactic, and cosmological) \\ \indent
\hspace{1.6cm} 98.35.Gi	Galactic halo \\ \indent
\hspace{1.6cm} 98.38.Jw	Infrared emission \\ \indent
\hspace{1.6cm} 98.56.-p	Local group; Magellanic Clouds

\maketitle

\vspace{-20cm}

\section{Introduction}

\hfill{\it The stone which the builders rejected,}

\hfill{\it  the same is become the head of the corner}

\hfill{Matthew, XXI:42}

\vspace{2mm}

The dark matter riddle was formulated for the Galaxy  in 1922 by Jacobus Kapteyn~\cite{Kapteyn}: there must be more mass than in stars,
the invisible or dark matter.  Currently it  is mostly approached from particle searches, aiming to find the weakly interacting massive particles (WIMPs),
  the purported cold dark matter (CDM) particle \cite{BertonePhysRep2005}.
A complementary approach is to investigate statistical properties of dark matter. Since the Galaxy exists billions of years, its dark matter, whatever its nature, 
probably had time enough to reach some type of equilibrium. This expectation is supported by the observed flattening of galactic rotation curves.
After all, a really constant circular rotation velocity $v_{\rm c}$ would imply, due to the Kepler relation $v_{\rm c}=[GM(r)/r\,]^{1/2}$, 
that the mass $M(r)$ inside a sphere of radius $r$ grows linearly in $r$, and hence that the mass density decays as $r^{-2}$. 
But that is the well known case of the singular isothermal sphere, a solution to the Poisson-Boltzmann equation, where the mass density
is a Boltzmann distribution, the equilibrium distribution of standard statistical physics.  The Galaxy is close to that.

The question ``Which type of dark matter causes the isothermal profile?'' has no established solution at present. But galaxy-scale difficulties for WIMP dark matter
are abundant,  see, e. g., \cite{Pawlowski2011}.  It is proper to approach the problem in the opposite direction:
 identify a candidate, and investigate whether it satisfies the relevant statistical criteria.
 
 The answer to this may  be related to the missing baryon problem: of all baryons expected from the standard model, 
 10\% are known to be in stars and 20-30\% in gas of galaxies and galaxy clusters, while the rest, 60-70\%, is still unaccounted for \cite{Gurzadyan}. 
 A new component, hot gas in the Galactic halo that weighs over 10 billion solar masses, 
 has been reported recently  \cite{Gupta2012}, but it will likely not be enough to solve the full problem.

Our argument starts with the remarkably low 15 K temperature of clouds all over the sky, known from IRAS observations,  and commonly attributed to dust, but see 
\cite{SGNC2012}. For lack of suitable instruments, 15 K objects would have mostly remained in the dark until recently, 
but by now  Planck detects 915 cold molecular cloud core candidates, out of over 15,000 unique sources that litter the sky at the 857 GHz
($350\mu$), the largest Planck frequency, see Fig. 6 of  \cite{PlanckVII}.
Several of these clouds have a gas mass  of 10,000 solar masses \cite{Juvela2010,Juvela2012}.
Towards the Galactic center, at coordinates $|b|<1^\circ$ and $\ell=300^\circ-330^\circ$, 1205 dark clouds are detected
by Herschel/Spitzer, which can be seen not only in absorption, but also in emission at $250\mu$ and $500\mu$ \cite{Wilcock2012}.

One particular  $15^\circ\times 12^\circ $ region on the sky  observed  already with B{\small{OOMERANG}} in the 1990's and focused on recently
 by Veneziani et al (2010) \cite{Veneziani2010},  exposes dozens of 7--20 K clouds. Their temperature, typical opening angle, two-dimensional 
 number density and emission intensity are explained from gravitational hydrodynamics (GHD) employing an isothermal sphere of Jeans clusters (JCs) 
 in which, next to stars and gas,  all dynamical mass is contained~\cite{NSG2011}.
It was concluded that {\it these clouds are JCs and make up the galactic dark matter and most of the missing baryons}.
This challenges  the standard \LCDM \ or concordance model, and clearly awaits testing by further observations.

Cold 15 K objects should be clearly exposed by Herschel SPIRE's 250$\mu$ detector.  
In early 2012 NASA/ESA published a multi-wavelength map of the Small Magellanic Clouds (SMC) \cite{NASASMC2012}.
 (A similar picture for the Large Magellanic Cloud \cite{NASALMC2012}  is  less suited for our purpose; it will be discussed in section 5.3.)
In the SMC picture one observes the bar on the right and the wing from the right to the middle, which both contain many stars.
Our interest is in the 250$\mu$ maps, represented as red dots in these color figures, that we reproduce for the SMC as gray dots in Figure 1.
Surprisingly, one observes many hundreds,  if not thousands, of HCCs all over the picture (except in regions without data), not exposed before in this number. 
Since most of them seem absent in the ``green'' map (160 and 100 $\mu$m), they are likely not background galaxies, but
must be cold ($\sim 15 $ K), and we adopt the Juvela (2012) name ``Herschel cold clouds'' (HCCs).  They are the focus of our present study. 
Our aim is to investigate whether the HCCs can correspond to JCs that would comprise the full dynamic dark matter of the Galaxy \cite{NSG2011}.

\begin{figure}
\hspace{0cm}
\includegraphics[width=15.7cm]{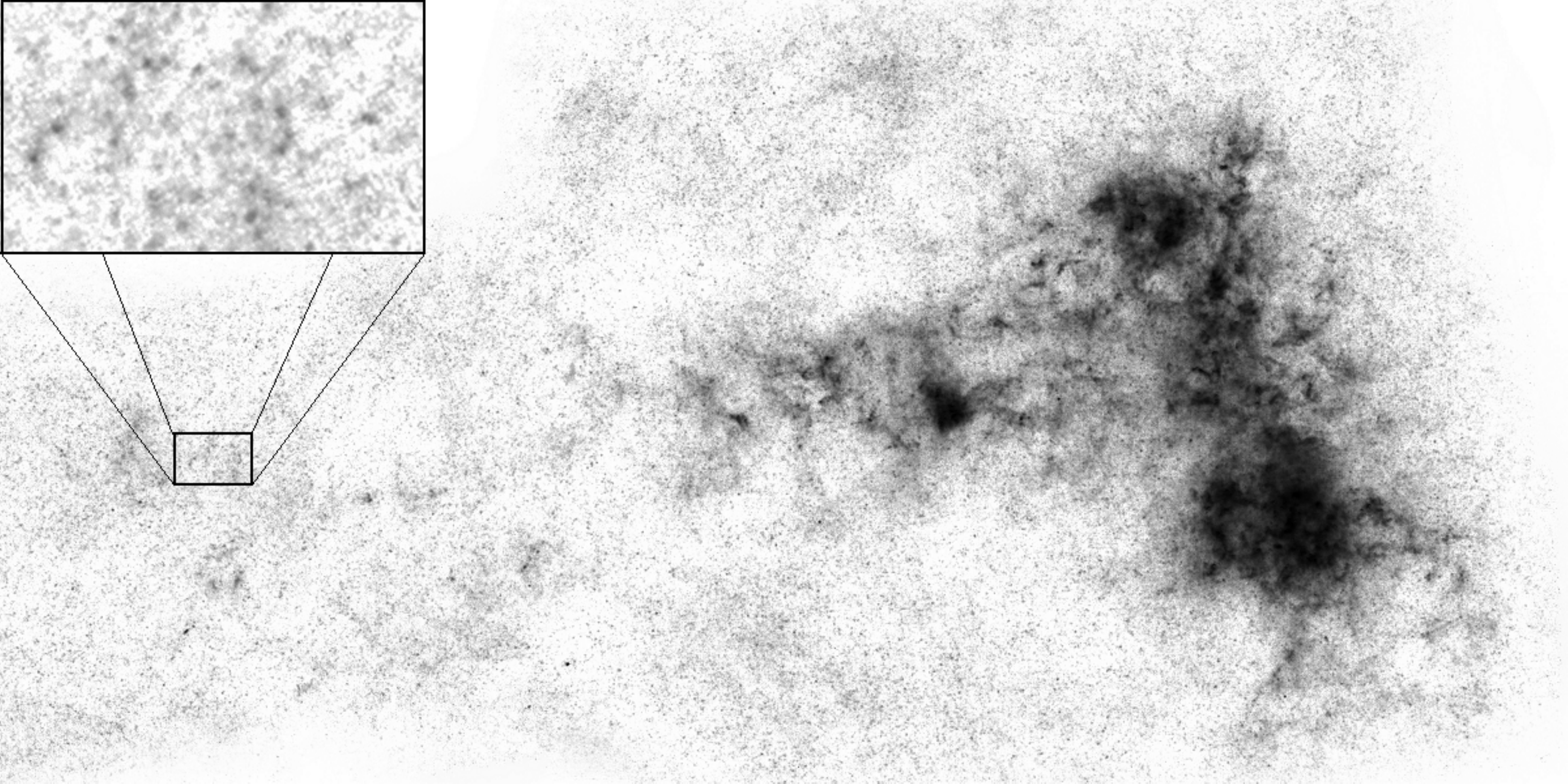} 
\caption{This luminosity-inverted $8.32^\circ\times 4.16^\circ$ Herschel-SPIRE map of the SMC at 250 $\mu$m exposes the star-rich bar on the right and 
the wing towards the left, and  thousands of Herschel cold clouds (HCCs) in the Galactic foreground, of different size and intensity.
This figure was prepared from data made public by NASA/ESA.}
\end{figure}

In section 2 we discuss the Jeans theory and its extension by gravitational hydrodynamics, and we recall a number of observations already 
explained by this model. In section 3 we formulate a model for the expected angular size distribution of the HCCs.
In section 4 we investigate their size distribution in the SMC map and in section 5 we discuss the fit to our model.
In section 6 we discuss the results within the view point of JCs and micro brown dwarfs (\mBDs), and we conclude with a discussion.

\section{The Jeans instability and GHD}

The Jeans 1902 mechanism asserts that when in the early Universe the plasma changes into a H-He gas,
this gas, considered to be infinitely distributed, exhibits instabilities at the Jeans scale, where pressure can no longer overcome the effects
of gravitation and diminished heat transfer~\cite{Jeans1902,Gibson1996}. 
This leads to the formation of Jeans gas clumps that weigh about 40,000 $M_\odot$.
Globular star clusters are supposed to have formed from a handful of coagulated clumps,  
but other ones are generally considered to have turned into stars, brown dwarfs, planets,  gas clouds, etc., so that they should not exist anymore, 
at least not in large numbers. We shall pose, however, the question: does the majority of the Jeans gas clumps still exist as HCCs?

In GHD it is asserted that, due to turbulence and viscosity, the Jeans clumps themselves fragment in
earth mass MACHOs, called  primordial gas particles \cite{Gibson1996} or, what we shall employ here, micro brown dwarfs (\mBDs) \cite{NSG2011}. 
This turned the Jeans gas clumps into JCs of
billions of  \mBDs. In order to reach a temperature of 2.725 K, the \mBDs \  would have to cool from gas to solid, 
by radiating  away the latent heat in order to cross the first-order phase transition line that emerges the H triple point at 13.8 K.
If this takes more than a Hubble time, their temperatures must become stuck somewhere above 14 K.

 The wide-binaries problem, the last-parsec problem  and the Helium-3 problem also find a simple explanation in GHD \cite{NHe3}, as do 
 long-duration radio transients~\cite{NSG2011,Kida2008,Frail2011}.
GHD offers plenty of dark baryons for making new stars even with modest recycling of stellar matter~\cite{Oort1970,Larson1972,newstars},
because it would consider the outflowing and infalling matter as dynamical behavior of the galaxy's large baryonic halo.
 The \mBDs \  would be leaky, so there is gas throughout the JC.
This offers an explanation for the Lyman-alpha forest: thousands of extinction lines in quasar spectra due to hydrogen  clouds along the sightline.
Why these clouds did not dissolve quickly is still a miracle \cite{Rauch1998}, but JCs of leaky \mBDs \ provide 
 a simple explanation: the gas is bound by mass concentrations, the \mBDs. 
 Therefore,  the total mass can be much larger than the gas mass, and this is the picture that we shall investigate in the present paper.

 The idea can also be tested in our immediate neighborhood. 
 The Local Leo Cold Cloud is  a large ($\sim5$ deg), cold (15-20 K), nearby (10-20 parsec) H cloud \cite{Peek2011}.
 Like the Lyman-alpha clouds, it can be long lived if it contains many unresolved \mBDs, which bind the gas gravitationally,
 thus explaining why the gas can have the observed large pressure discussed in \cite{MeyerPeek2012}.
 The observed velocity differences inside the Local Leo Cold Cloud are explained by the local winds due to mostly  random motion of \mBDs, 
 that sweep the gas with them.

In GHD, planet and star formation occurs by coalescence of \mBDs ~\cite{NSG2011,SchildGibson2011,GibsonSchild2011}.
They are cold and relatively small now (solar size),  but when heated they expand and can coagulate. 
First a gas planet is formed and, after enough steps, a star is created. This explains why
galaxy mergers such as Antennae \cite{Fall2005}  and Tadpole \cite{Tran2003}
have tails of thousands of young ($<10$ Myr) globular star clusters, much younger than the event duration of galaxy merging (150-200 Myr):
during the merging, the local JCs of one galaxy are heated by 
the center of the other, so that they later develop stars~\cite{NSG2011}. 
Many or possibly all stars may have been formed in such clusters ~\cite{Fall2005}. 

\subsection{The Jeans mass}

We shall need a mechanism that explains the formation of so many, so regular, so identical cold clouds. 
A well known candidate is the 1902 Jeans instability of the neutral gas, just as it forms at the recombination. 
It is often taught as a separate issue in standard courses on astrophysics, but for us it will be a cornerstone.

The Jeans mechanism asserts that at last scattering  ($L$; decoupling, recombination) the plasma turns into a neutral gas, which is unstable 
on a scale where pressure and heat transport are overcome by gravitation. The free fall time is $\tau_g =(G\rho_B)^{-1/2}= 5.30$ Myr,
where the baryonic density at the decoupling at $z_L=1088$ is  $\rho_B=\Omega_B\rho_c(1+z_L)^3=5.35 \,10^{-22}$ gr cm$^{-3}$ for $\Omega_B=0.045$ and 
$\rho_c=9.21\,10^{-30}$ gr cm$^{-3}$ at $h_{70}=1$. The speed of sound of a monoatomic gas is $v_s=({5 p/3\rho})^{1/2}$. 
For the gas of H and 25\% of weight in He the pressure $p=0.813\rho k_BT/m_N$ yields $v_s=5.68$ km s$^{-1}$. 
The gas fragments at the Jeans scale $L_{\rm Jeans}=v_s\tau_{g}=31$ pc into Jeans clumps with mass

\BEQ 
\label{Mjeans} \hspace{-1.cm}
M_{\rm Jeans}=\frac{\pi}{6}\rho_{B}L_{\rm Jeans}^3=  \frac{\pi}{6}\Omega_B\rho_c(1+z_L)^3L_{\rm Jeans}^3 = \frac{\pi v_s^3}{6G^{3/2}\rho_B^{1/2}}= 40,000\,M_\odot.\EEQ
A practical way to check this is to consider globular star clusters,
which, we presume, are the result of up to a dozen or so of JC mergers, as is exhibited by their sets of stars of different ages \cite{Puzia-Brodie-1999}.
To yield a GC mass between ten thousand and a few hundred thousand solar masses,  the Jeans mass should indeed be a few ten thousand solar masses.

\subsection{Nearly isothermal mass density profile}

The circular rotation speed at the position of the Sun, taken as $\vsun =220$ km s$^{-1}$, drops within 10\% accuracy mildly to 
175 km s$^{-1}$ at 60 kpc \cite{Xue2008}. This range can be modeled as $v_c(r)=\vsun (d_{\rm SA}/r)^\varepsilon$ km s$^{-1}$ with $\varepsilon=0.114$ and 
$d_{\rm SA}=8.0\pm0.5$ kpc, the distance to \SagA. Now it is found that $\vsun=240\pm 14$ km s$^{-1}$ \cite{Honma2012}.
In the presence of spherical symmetry, Kepler's law $GM(r)r^{-1}=v_c^2(r)$ then yields for the mass density $\rho=M'(r)/4\pi r^2$:


\BEQ \rho(r)=\frac{(1-2\varepsilon)\vsun ^2}{4\pi G\, r^2}\left(\frac{d_\SA}{r}\right)^{2\eps}, \qquad 
\vsun =240\pm14 \textrm{ km s}^{-1}.
\EEQ
For $\eps=0$ this is the singular isothermal sphere model with velocity dispersion $\vsun /\sqrt{2}$.

\section{Model for the angular size distribution of Galactic foregrounds}

We assume that the HCCs make up a fixed fraction $c_\hcc\le1$  of the total mass density. With average size $\barR$ and average mass $\bar M$,
this leads us to their average number density  

\BEQ \hspace{-1cm}
n(r)=\frac{c_\hcc}{\barM}\rho(r)\equiv \frac{\bA}{\bar R\, r^{2}}\left(\frac{d_\SA}{r}\right)^{2\eps},\qquad 
\bA=\frac{(1-2\eps)\bar R\vsun ^2c_\hcc}{4\pi G\bar M}.
\EEQ
For the distribution of radii $R$ at position ${\bf r}$ we consider the factorized form $n(r,R)$ = $n(r)f(R/\bar R)/\bar R$ with 
$\int_0^\infty\d x \,f(x)=\int_0^\infty\d x \,xf(x)=1$, which reproduces the local  average $\int\d R\,n(r,R)=n(r)$, while evidently $\langle R\rangle=\barR$.
We specify to the  $\Gamma$-distributions $f(x)=(nx)^n\exp(-nx)/{x\Gamma(n)}$, which for $n\to\infty$ lead to
the fixed-$R$ case $f(x)=\delta(x-1)$.

\subsection{Seen from SagA$^\ast$}

From SagA$^\ast$, an HCC of physical radius $R$ is seen with angular radius $\theta=\arctan R/r\approx R/r$ and  angular diameter $2\theta$. 
Let $N(\theta)$ denote the number of HCCs that have an angular radius smaller than $\theta$.
Its differential increase per unit angular radius per unit solid angle is 

\BEQ
\hspace{-0cm} 
\frac{\dtwee N
(\theta)}{{\rm d} \theta\d\Omega}=\frac{1}{\Delta \Omega}\int \hspace{-2mm}\int \d r\d\Omega\d R \,r^2 
n(r,R)\delta\left(\theta-\frac{R}{r}\right), 
\EEQ
where $\d\Omega$ is the differential solid angle,
the integral of which cancels the factor $\Delta\Omega$.
Using $\delta(\theta-Rr^{-1})=R\theta^{-2}\delta(r-R\theta^{-1})$ and defining $x=R/\bar R$, the result takes the form


\BEQ\label{thdgdthSA}
\hspace{-2.cm}
\frac{\dtwee N
(\theta)}{\d \theta\d\Omega}=\frac{\bA \bC(\theta)}{\theta^2},\hspace{4mm}
\bC(\theta)=\int_0^\infty \d x f(x)x  \left(\frac{d_\SA\theta}{x\bar R}\right)^{2\eps}
=\frac{\Gamma(n+1-2\eps)}{\Gamma(n+1)}  \left(\frac{nd_\SA\theta}{\bar R}\right)^{2\eps},\EEQ
which has  dimension rad$^{-3}$.  Notice that $\bA$ and $\bar R$ do not occur independently here. 
For $\eps\to0$ one considers in fact isothermally distributed HCCs, 
 where this relation has $\bC=1$ and expresses the isothermal relation $\rho(r)\sim r^{-2}$  in a dimensionless form.

\subsection{Seen from the earth}

The fact that the Sun is not located in the center of the Galaxy  introduces a slight complication.
But actually it turns out as a benefit that allows us to estimate both the amplitude $A$ and the average radius $\bar R$ from the statistics of the data.

A sphere of radius $R=x\bar R$ at distance $d$ from the Earth at galactic coordinates ($b,\ell$) has position 
${\bf  r}=(d\cos b\cos\ell+\DSA , d\cos b\sin \ell,d\sin b)$ with respect to the center of the Galaxy. 
The modulus is $ r(d,b,\ell)=(d^2+2\DSA d\cos b \cos \ell+\DSA ^2)^{1/2}$
and the volume element  $\d^3r=\d d\d\Omega \,d^2$, where $\d\Omega=\d b\d\ell\,\cos b$. Because $\theta=R/d$  we have the definition

\BEQ
\frac{\dtwee N(\theta)}{\d \theta\d\Omega}=\frac{1}{\Delta\Omega} \int \hspace{-2mm}\int\d d\,\d \Omega\, \d x \, d^2 n(r)f(x)\delta\left(\theta-\frac{x\bar R}{d}\right).
\EEQ
When $\Delta\Omega$ is a small field around ($\bar b,\bar\ell$), the integrand hardly varies with $\Omega$, so 
 $\Delta\Omega$ drops out again.
The $d$-integral yields 
${\dtwee N(\theta)}/{\d \theta\d\Omega}=\bA \theta^{-2}\bC(\theta)$, with 
 $\bA$ from (3) and now




\BEQ \hspace{-0cm}
\label{CSun=}
\bC(\theta)=\int_0^\infty \d x f(x)x\,\frac{({x\bar R}{\theta}^{-1})^2d_\SA^{2\eps}}{[r({x\bar R}{\theta}^{-1},\bar b,\bar \ell)]^{2+2\eps}}.
\EEQ

\section{Data analysis of Herschel maps of the Magellanic Clouds}

\subsection{Detection}
The Herschel ``Spectral and Photometric Imaging Receiver'' (SPIRE)
instrument has detectors for central wavelengths of $250\mu$, $360\mu$ and $520\mu$. 
The full-width at half-maximum  of the 250$\mu$ detector is $18.1''$, see section 6.4 of Ref. \cite{Herschel-SPIRE}.
Herschel-SPIRE has delivered a multi-frequency map of the SMC  \cite{NASASMC2012}.
The  250$\mu$ map  has a resolution of  $6000\times 3000$ pixels and its angular size
is determined by identifying a number of spots with stars known from catalogs. 
We thus obtain $8.315^\circ\times 4.157^\circ$, which corresponds to  pixels of linear size of  $\theta_\pix=4.99''$.
Hence the resolution limit is 3.6 pixels across.

We shall present data for structures from 8 to 124 pixels in size.
Each pixel has a solid angle $\theta_\pix^2$. 
Approximating an object with a number of pixels $N_\pix$ as a sphere, we define the angular radius $\theta$ by
$N_\pix\theta_\pix^2=\pi\theta^2$.
In the analysis the angular  widths are in the range $15'' < 2\theta < 63''$, the smallest value being just below the resolution limit,
but not causing a clear deviation in Fig. 2.

\subsection{On ImageJ}

To count the HCC's we use the the freely available program ``ImageJ''. 
We first split the colors of the multi-frequency map to extract the 250$\mu$ map represented in  Fig. 1.
Next we mention some relevant points/decisions made with respect to the counting method.

1)	First of all, we set the minimum threshold value to 11 and the maximum to 255 (on a 256 bit gray-scale) to select the pixels with a grayscale between 11 and 255.  
If the minimum threshold value is set much higher, visible absence of data will occur; setting the threshold much lower than 11 causes too much background noise. 
Here we define background noise as very faint objects measuring less than $2\times2$ pixels, which is below the diffraction limit of 
$\sim4\times4$ pixels for Herschel's 3.5 meter primary mirror at this wavelength.

2) We use a watershed function to split overlapping HCCs. 
This also helps the edge tracker (see point 5) to distinguish between the background and HCCs.  

3) We count the number $\Delta N$ of HCCs for pixel sizes between 6 and 125.

4) Large distortions (defined as HCCs bigger than  $\sim130$ pixels), like luminous matter clumps,
 are unusable. They occupy 45-55\% of the map, while some 12\% of the map is without any data. In such regions we normalize by the analyzed area, see 6).
 
5)	For the counting we use an edge tracker to help distinguishing nearby/overlapping HCCs. However, it seems that the edge tracker increases the apparent area size
 of HCCs as a function of their true area size. We correct for this issue in our data by the empirical relation $N_\pix^{\rm true}=(0.11\pm0.04) (N_\pix^{\rm apparent})^{1.41\pm0.01}$.

6) We bin the data  in sets of $\Delta N_\pix=4$ pixels, viz. $N_\pix=(6,7,8,9)$, ($10,11,12,13$), etc., with central 
bin values $\bar N_\pix=7.5,\, 11.5$ etc., to be used in the definition of $\theta$.
We determine $\theta^2\dtwee N/\d \theta\d\Omega\equiv 2\pi^{-1/2} \bar N_\pix^{3/2}\theta_\pix\cdot\Delta N/\Delta N_\pix\Delta\Omega$, 
where $\Delta N$ is the number of HCCs in the bin of the investigated region of angular area 
$\Delta\Omega=\cos\bar  b\Delta b\Delta \ell$.

7) We determine the error bars by measuring 7 different areas in our data set and calculating the standard deviation for each bin of 4 pixels.
 
\subsection{Two ways of analysis for the SMC}

We divide the SMC map into 7 ``good''  regions that are poor in stars and define region 8 as the rest, centering on the star-rich regions, on
 the border regions without data, and on other regions with poor data.
 The average angular position is $(\bar b,\bar\ell) =(-44.2^\circ, 302.8^\circ)$. 


In our first method, we analyze the full SMC map. ImageJ treats the luminous regions as big HCCs of 10,000 pixels and more, so they do not add to our counts.
We correct the useful area by subtracting these big luminous regions, as well as regions of HCCs with more than 130 pixels.
We do the same for the side regions without data. The error in the variable $y\equiv\theta^2\dtwee N(\theta)/\d\theta\d\Omega$ is defined as
$\Delta y=[\sum_{i=1}^nw_i(y_i-\bar y)^2\sum_{i=1}^nw_i^2]^{1/2}$, where $n=8$ and $w_i=\Delta\Omega_i/\sum_{j=1}^n\Delta\Omega_j$
is the areal weight of the region $i$ and $\bar y=\sum w_iy_i$.

In our second method for the SMC, we select the seven ``good'' areas away from the star-rich regions, we analyze them and we join them by area size.
The error in the variable $y$ then sums over these $n=7$ areas.

For the LMC only a strip on the left and a smaller strip on the right are useful for our analysis. The errors are now larger, since
we only have two regions ($n=2$) with relatively few HCCs. The average position of the LMC is $(\bar b,\bar\ell) =(-32.89^\circ, 280.47^\circ)$.

 \subsection{Provisos of our method}

We take for granted that the ``red dots'' of the color version of the SMC map (the gray ones in Fig. 1)
represent physical objects and we model them as spheres. 
One may argue that the observations present more edgy or hairy objects, which could be an effect of a mutual interactions, near passages or
central collisions, or complete absence of nearly spherical objects. But most of these issues concern details below the resolution limit, which cannot
 be analyzed properly. In the case of large objects, we assume that they are agglomerations of spheres, and separate them accordingly.
But there can be mass streams between them, in particular due to mutual  interactions.
However, given the size of our statistical errors, these issues do not seem too relevant at our level of description.

The LMC map \cite{NASALMC2012} is largely dominated by the star-rich region. For our analysis small side regions remain available,  the
results of which agree with those for the SMC.

Although we consider our model as reasonable and the parameters come out as reasonable, we stress that the data could be fit to many other models.

We have analyzed only the MC regions, which cover a small piece of the sky.

In this study we did not have access to the infrared intensities, which give additional information about the ratio of their radiating surface
to their distance squared. 

\newcommand{\pmt}{$\pm$}

\myskip{ 1) FixedR SMCtot
23.6105 chisqIs
0.843232 chisqpnuIs
Aepsis (102.304 + 3.62477 pm)
myRis pc (1.68276 + 0.179999 pm)
AngleSMCis arcsec (5.69007 + 0.608648 pm)
AngleLMCis arcsec (7.15658 + 0.765516 pm)
MassIs MMo (13.5459 + 1.86919 pm)
kmPerSec (4.16095 + 0.0737137 pm) SpeedIs}

\newcommand{\fSMCtotXi}{23.6}
\newcommand{\fSMCtotXinu}{0.84}
\newcommand{\fSMCtotA}{102\pmt4}
\newcommand{\fSMCtotR}{1.68\pmt0.18}
\newcommand{\fSMCtotth}{5.7\pmt 0.6}
\newcommand{\fSMCtotM}{13.5\pmt1.9}
\newcommand{\fSMCtotsig}{4.16\pmt0.07}

\myskip{ 2 FixedR SMCgood
67.1088 chisqIs
2.39674 chisqpnuIs
Aepsis (106.304 + 3.38111 pm)
myRis pc (1.81887 + 0.183679 pm)
AngleSMCis arcsec (6.15032 + 0.621089 pm)
AngleLMCis arcsec (7.73545 + 0.781163 pm)
MassIs MMo (14.0908 + 1.81667 pm)
kmPerSec (4.08192 + 0.0649148 pm) SpeedIs}

\newcommand{\fSMCgoodXi}{67.1}
\newcommand{\fSMCgoodXinu}{2.40}
\newcommand{\fSMCgoodA}{106\pmt3}
\newcommand{\fSMCgoodR}{1.82\pmt0.18}
\newcommand{\fSMCgoodth}{6.2\pmt 0.6}
\newcommand{\fSMCgoodM}{14.1\pmt1.8}
\newcommand{\fSMCgoodsig}{4.08\pmt0.06}

\myskip{3 FixedR LMCsides
12.4609 chisqIs
0.445031 chisqpnuIs
Aepsis (112.519 + 14.3896 pm)
myRis pc (1.17238 + 0.191985 pm)
AngleSMCis arcsec (3.96427 + 0.649176 pm)
AngleLMCis arcsec (4.98599 + 0.81649 pm)
MassIs MMo (8.58071 + 2.44546 pm)
kmPerSec (3.96758 + 0.253699 pm) SpeedIs}

\newcommand{\fLMCsideXi}{12.5}
\newcommand{\fLMCsideXinu}{0.45}
\newcommand{\fLMCsideA}{113\pmt14}
\newcommand{\fLMCsideR}{1.17\pmt0.19}
\newcommand{\fLMCsideth}{5.0\pmt0.8}
\newcommand{\fLMCsideM}{8.6\pmt2.4}
\newcommand{\fLMCsidesig}{3.97\pmt0.25}

\myskip{4 FixedR SplusLMC
38.1238 chisqIs
0.657307 chisqpnuIs
Aepsis (103.586 + 3.45599 pm)
myRis pc (1.64983 + 0.148205 pm)
AngleSMCis arcsec (5.57872 + 0.501137 pm)
AngleLMCis arcsec (7.01654 + 0.630296 pm)
MassIs MMo (13.1166 + 1.56309 pm)
kmPerSec (4.13513 + 0.0689812 pm) SpeedIs}

\newcommand{\fSMCLMCXi}{38.1}
\newcommand{\fSMCLMCXinu}{0.66}
\newcommand{\fSMCLMCA}{104\pmt3}
\newcommand{\fSMCLMCR}{1.65\pmt0.15}
\newcommand{\fSMCLMCth}{5.6\pmt0.5}
\newcommand{\fSMCLMCM}{13.1\pmt1.6}
\newcommand{\fSMCLMCsig}{4.14\pmt0.07}

\myskip{5 RandomR SMCtot
22.3794 chisqIs
0.799265 chisqpnuIs
Aepsis (109.761 + 4.56532 pm)
myRis pc (1.30438 + 0.177788 pm)
AngleSMCis arcsec (4.41063 + 0.601169 pm)
AngleLMCis arcsec (5.54739 + 0.75611 pm)
MassIs MMo (9.78674 + 1.70468 pm)
kmPerSec (4.01712 + 0.0835424 pm) SpeedIs}

\newcommand{\rSMCtotXi}{22.4}
\newcommand{\rSMCtotXinu}{0.80}
\newcommand{\rSMCtotA}{110\pmt5}
\newcommand{\rSMCtotR}{1.30\pmt0.18}
\newcommand{\rSMCtotth}{4.4\pmt 0.6}
\newcommand{\rSMCtotM}{9.8\pmt1.7}
\newcommand{\rSMCtotsig}{4.02\pmt0.08}

\myskip{6  RandomR SMCgood
65.5018 chisqIs
2.33935 chisqpnuIs
Aepsis (113.694 + 4.17081 pm)
myRis pc (1.43053 + 0.181738 pm)
AngleSMCis arcsec (4.83717 + 0.614526 pm)
AngleLMCis arcsec (6.08386 + 0.772908 pm)
MassIs MMo (10.3619 + 1.66207 pm)
kmPerSec (3.94703 + 0.0723973 pm) SpeedIs}

\newcommand{\rSMCgoodXi}{65.5}
\newcommand{\rSMCgoodXinu}{2.34}
\newcommand{\rSMCgoodA}{114\pmt4}
\newcommand{\rSMCgoodR}{1.43\pmt0.18}
\newcommand{\rSMCgoodth}{4.8\pmt 0.6}
\newcommand{\rSMCgoodM}{10.4\pmt1.7}
\newcommand{\rSMCgoodsig}{3.95\pmt0.07}

\myskip{ 7 LMCsides RandomR
13.3567 chisqIs
0.477026 chisqpnuIs
Aepsis (126.926 + 19.8485 pm)
myRis pc (0.793044 + 0.179863 pm)
AngleSMCis arcsec (2.68159 + 0.608187 pm)
AngleLMCis arcsec (3.37272 + 0.764936 pm)
MassIs MMo (5.14551 + 1.9426 pm)
kmPerSec (3.73563 + 0.292086 pm) SpeedIs}

\newcommand{\rLMCsideXi}{13.4}
\newcommand{\rLMCsideXinu}{0.48}
\newcommand{\rLMCsideA}{127\pmt20}
\newcommand{\rLMCsideR}{0.79\pmt0.18}
\newcommand{\rLMCsideth}{3.4\pmt0.8}
\newcommand{\rLMCsideM}{5.2\pmt1.9}
\newcommand{\rLMCsidesig}{3.74\pmt0.29}

\myskip{8 RandomR SplusLMC
37.4851 chisqIs
0.646294 chisqpnuIs
Aepsis (111.6 + 4.39479 pm)
myRis pc (1.253 + 0.144206 pm)
AngleSMCis arcsec (4.23687 + 0.487617 pm)
AngleLMCis arcsec (5.32885 + 0.613292 pm)
MassIs MMo (9.24628 + 1.39707 pm)
kmPerSec (3.98389 + 0.0784423 pm) SpeedIs}

\newcommand{\rSMCLMCXi}{37.5}
\newcommand{\rSMCLMCXinu}{0.65}
\newcommand{\rSMCLMCA}{112\pmt4}
\newcommand{\rSMCLMCR}{1.25\pmt0.14}
\newcommand{\rSMCLMCth}{4.2\pmt0.5}
\newcommand{\rSMCLMCM}{9.2\pmt1.4}
\newcommand{\rSMCLMCsig}{3.98\pmt0.08}

\begin{table}
\hspace{-5mm}
\renewcommand{\arraystretch}{1.1}
\begin{tabular}{|l|l |r| r|  l|  l|  r|  r|r|}
\hline \hline
Region & $n$ &$\chi^2$&$\chi^2/\nu$& $\bA$ &$\bar R$(pc)&$\bar\theta_{\rm MC}('')$&$\bar M(10^3M_\odot)$&$\sigma_\bd$(km/s)\\ \hline\hline
SMC-total& $\infty$& \fSMCtotXi& \fSMCtotXinu& \fSMCtotA& \fSMCtotR& \fSMCtotth& \fSMCtotM& \fSMCtotsig \\ \hline
SMC-good& $\infty$& \fSMCgoodXi& \fSMCgoodXinu& \fSMCgoodA& \fSMCgoodR& \fSMCgoodth& \fSMCgoodM& \fSMCgoodsig \\ \hline
LMC-sides& $\infty$& \fLMCsideXi& \fLMCsideXinu& \fLMCsideA& \fLMCsideR& \fLMCsideth& \fLMCsideM& \fLMCsidesig \\ \hline
S+L MC& $\infty$& \fSMCLMCXi& \fSMCLMCXinu& \fSMCLMCA& \fSMCLMCR& \fSMCLMCth& \fSMCLMCM& \fSMCLMCsig \\ \hline \hline
SMC-total& $2$& \rSMCtotXi& \rSMCtotXinu& \rSMCtotA& \rSMCtotR& \rSMCtotth& \rSMCtotM& \rSMCtotsig \\ \hline
SMC-good& $2$& \rSMCgoodXi& \rSMCgoodXinu& \rSMCgoodA& \rSMCgoodR& \rSMCgoodth& \rSMCgoodM& \rSMCgoodsig \\ \hline
LMC-sides& $2$& \rLMCsideXi& \rLMCsideXinu& \rLMCsideA& \rLMCsideR& \rLMCsideth& \rLMCsideM& \rLMCsidesig \\ \hline
S+L MC& $2$& \rSMCLMCXi& \rSMCLMCXinu& \rSMCLMCA& \rSMCLMCR& \rSMCLMCth& \rSMCLMCM& \rSMCLMCsig \\
\hline\hline
\end{tabular}
\renewcommand{\arraystretch}{1.0}
\caption{Fits of the model for Jeans clusters
for the four data sets: the total SMC, the good regions of the SMC, the sides of the LMC, and the total SMC combined with the sides of the LMC. 
$\bA$ is an amplitude; $\bar R$ the average radius of the HCC;  $\bar\theta_{\rm MC}$ refers to $\bar\theta_\SMC$, the angular radius of
an average HCC located at the SMC, except for the LMC-sides cases,
where it correspondingly refers to the LMC. Finally, $\bar M$ is the average HCC mass needed to explain all dynamical mass in the Galaxy,
and $\sigma_\bd$ is the velocity dispersion of the \mBDs.
The upper data refer to HCCs with a fixed radius ($n=\infty$) and the lower to
radii taken from the  $n=2$ $\Gamma$-distribution $f(x)=4x\exp(-2x)$.}
\end{table}

\section{Fits to the nearly isothermal model}
 \subsection{Towards the SMC, the case of equal radii}

A $\chi^2$ fit of the theory of previous section to the observed 30 binned values of $\theta^2\d N/\d \theta\d\Omega$ has been performed, see Fig. 2.
For the full SMC picture the fit yields

\renewcommand{\pmt}{\pm}
\myskip{
\newcommand{\fSMCtotXi}{23.7}
\newcommand{\fSMCtotXinu}{0.85}
\newcommand{\fSMCtotA}{103\pmt4}
\newcommand{\fSMCtotR}{1.73\pmt0.18}
\newcommand{\fSMCtotth}{5.9\pmt 0.6}
\newcommand{\fSMCtotM}{11.6\pmt1.6}
\newcommand{\fSMCtotsig}{3.79\pmt0.07}
}

\BEQ
\label{SMCresultfixR}
\hspace{-1cm}  \bA=\fSMCtotA, \qquad \bar R=\fSMCtotR \textrm{ pc},  \qquad  \chi^2=\fSMCtotXi,  \quad\frac{\chi^2}{\nu}=\fSMCtotXinu
\EEQ
where $\nu=30-2$ is the number of degrees of freedom.
The error bars derive from the inverse correlation matrix of the fit parameters $p_1\equiv \bA$ and $p_2\equiv \bar R$pc$^{-1}$ for the prediction
$y_i\equiv A\,C(\theta_i;\bar R)$, viz. $C^{-1}_{ab}=2\sum_i(\p y_i/\p {p_a})(\p y_i /\p {p_b})\sigma_i^{-2}$.
It follows that $\langle \delta A^2\rangle\equiv C_{11}=
22.0
$, $\langle \delta A \delta \bar R\rangle $pc$^{-1}\equiv C_{12}=
-0.71
$ and  $\langle\delta\bar R^2\rangle $pc$^{-2}\equiv C_{22}=
0.032$.

The angular radius of a HCC of average size, located in the SMC, is $\bar\theta_{\rm SMC}=\bar R /d_{\rm SMC}$,
with $d_{\rm SMC}=61\pm3$ kpc the SMC distance.  Hence $\bar\theta_{\rm SMC}=(\fSMCtotth)''$,  which is indicated in Fig. 2.
(By linear regression, we express here and below all errors in terms of $C_{ab}$.)

To test whether the HCCs can belong mainly to the SMC, we follow our second method and discard the data from the star-rich SMC region, about  $51\%$ of all data.
But this appears to yield only minor shifts in $\bA$ and $\bar R$, see the second row in Table 1.

Both the small values of $\bar\theta_\SMC$ and the minor shifts with respect to leaving out star-rich regions
reject the assumption that the HCCs of Fig. 1 belong to the SMC.
Hence we conclude that they nearly all belong to the Galactic Halo. 

The investigated distances stem from $\bar R/\theta$ and lie between 10 and 40 kpc.
All fit parameters are tabulated in Table 1. As a canonical average radius,  we adopt $\bar R=2$ pc.

 \subsection{Towards the SMC, spread in radii}
 
 \noindent

The radii of the HCCs can be random, mixing angular sizes of large, far HCCs with nearby, small ones. 
 As an example, we consider the $n=2$ $ \Gamma$-distribution $f(x)$ $=$ $4x\exp(-2x)$.
Rows 5, 6 and 8 in Table 1 show differences with respect to the equal-$R$ cases (rows 1,2 and 4, respectively).
These modest deviations make it unlikely that the size distribution can play an important role in the physical 
interpretation of the results.

\myskip{
, both for the total SMC and 

 $\chi^2=21.8=0.78\nu$ and

\BEQ
\label{SMCresultSal}
&&  \hspace{-2cm} 
A=164\pm 13,  \quad \bar R=1.69\pm 0.37  \textrm{ pc}, \quad \bar\theta_{\rm SMC}=5.7''\pm 1.3''.
\EEQ
using $\quad \{C_{11},C_{12},C_{22}\}=\{168,-4.65, 0.137\}$.
When we again leave out the data from the star-rich SMC region, the fit yields $\chi^2=64.7=2.31\nu$, $A=170\pm11$, $\bar R=1.89\pm0.36$ pc, 
and $\bar\theta_{\rm SMC}=6.4''\pm1.2''$. The modest deviations  from Eq. (\ref{SMCresultfixR}) 
make it thus unlikely that the size distribution plays an important role in the physical meaning of the results.
}

\begin{figure} \label{data+fit}
\centerline{\includegraphics[width=10cm]{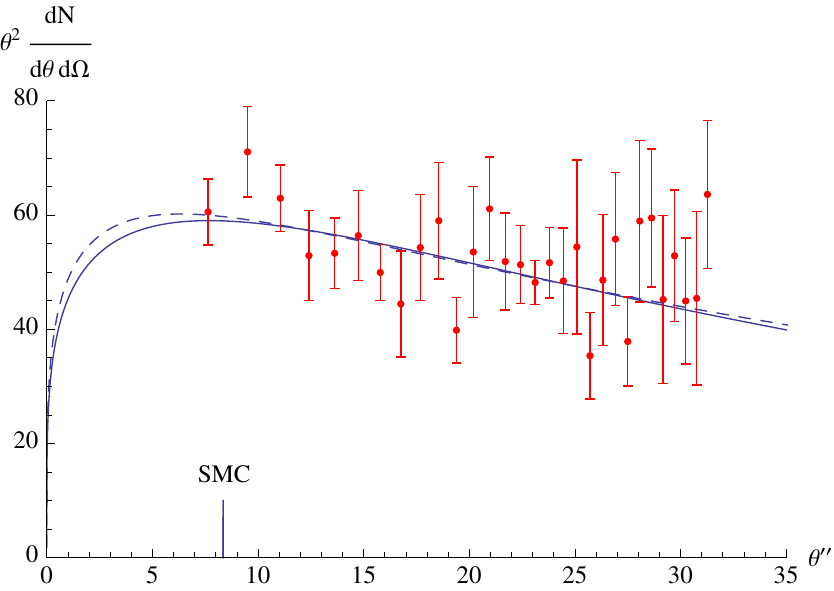}}
\caption{$N(\theta)$ is the number of Herschel cold clouds (HCCs) with angular radius smaller than $\theta$. 
The data for $\theta^2\dtwee N/\d\theta\d\Omega$  (in rad$^{-1}$) as function of $\theta$ (in arcsec) are deduced from Fig. 1.
The full curve is the best fit for HCCs of equal radius and the dashed curve for HCC radii taken from a $\Gamma$-distribution with $n=2$.
The initial growth stems from the $\theta^{2\eps}$ factor and the decay from the power of  $dr^{-1}$  in Eq. (7). 
The vertical bar denotes the angular radius $5.7''$ of an HCC of average radius located in the SMC.}
\end{figure}

\subsection{Towards the LMC}

A similar map exists for the LMC \cite{NASALMC2012} and in the data acquisition, described above, 
we only use two side regions.
Now the error bars are much larger, which actually leads to a small $\chi^2$ (Table 1, rows 3 and 7).
We consider the deviations from the fit parameters for the SMC (Table 1, rows 1, 2 and 5, 6) to be statistically unimportant.

\myskip{
For fixed radii the fit now yields a small $\chi^2=13.4$ due to the large error bars, and 

\BEQ
\label{LMCresultSal}
  \hspace{-1.5cm} 
&A=204\pm 33,  \quad \bar R=1.69\pm 0.40  \textrm{ pc}, \quad \bar\theta_{\rm SMC}=5.7''\pm 1.4'', &\textrm{    fixed $R$}, 
\EEQ
with $\{C_{11},C_{12},C_{22}\}=\{1076,-12.3, 0.16\}$. 
Within two standard deviations this overlaps with the SMC results. The data are too noisy for a meaningful random-$R$ fit.
}

\subsection{SMC combined with LMC}

Combining the 30 data for the total SMC with the 30 for the  LMC, hardly changes the parameters of the total SMC alone
and decreases only slightly the error bars (Compare rows 5 and 8 in table 1 with rows 1 and 4, respectively). 
This shows again that the LMC data, with their larger error bars,  have low statistical weight.

\section{Interpretation in terms of Jeans clusters and micro brown dwarfs}

Recalling Eq. (3), we get for the average mass of the HCCs from the SMC

\BEQ\label{Mav=}
\hspace{-2.2cm}
\frac{\bar M}{c_\hcc}=\frac{\bar R\vsun ^2(1-2\eps)}{4\pi G\bA}=(\fSMCtotM)10^3M_\odot \textrm{  or  } (\fSMCgoodM) 10^3M_\odot. \EEQ
The first (second) estimate holds the assumption of fixed radius for the full (restricted) data set; 
for the case of random $R$ this becomes $(\rSMCtotM)10^3M_\odot$ or ($\rSMCgoodM)\times 10^3M_\odot$ for these sets, respectively.
For the LMC the results deviate, see Table 1, but, as discussed, they have lower statistical weight.

Without further input, $c_\hcc$ may take a value of order $10^{-3}-10^{-4}$, the case of solar mass gas clouds, as often concluded from gas masses alone.
However, if the clouds are composed of leaky \mBDs, the total mass  can be much larger.

The Jeans mass has been estimated as $40,000M_\odot$ in Eq. (1). 
Therefore, it is imperative to discuss the case $c_\hcc=1$, for which the HCCs represent the complete dynamical dark matter of the Galaxy.
One then identifies the HCCs of heretofore unknown origin with the JCs predicted by the Jeans instability and the GHD fragmentation into \mBDs.
In view of table 1,  we adopt the canonical average mass $\bar M_\jc=10,000M_\odot$.
This lies within the range of HCC masses deduced  from gas alone \cite{Juvela2010,Juvela2012,Kainulainen2011}.
The typical distance to the nearest JC is  $d_1=[\rho(d_\SA)/\bar M_\jc]^{-1/3}=92$ pc,  its angular radius being $1.1^\circ$.  
Cold structures of size $\sim0.5^\circ$ have been discussed in \cite{Veneziani2010},
 while the HCCs analyzed in \cite{Juvela2010,Juvela2012} have distances between 225 pc and 1.67 kpc.

If the JCs themselves are isothermal distributions of \mBDs \ with velocity dispersion $\sigma_\bd$, their mass and radius are related as
$GM_\jc=2\sigma_\bd^2R_\jc$.  Assuming that $\sigma_\bd$ is basically  a constant,  this relation also holds for the average.
We can thus express $\bA$ from (3) with $c_\hcc=1$ as $\bA=(1-2\eps)\vsun ^2/8\pi\sigma_\bd^2$, which yields, for fixed-$R$,
$\sigma_\bd=\fSMCtotsig$ km s$^{-1}$ $(\fSMCgoodsig$ km s$^{-1})$ for the first (second, restricted) SMC data set, and
when modeled with random radii $\sigma_\bd=\rSMCtotsig$ km s$^{-1}$  ($\rSMCgoodsig$ km s$^{-1}$), 
respectively. The LMC yields $\fLMCsidesig$ km  s$^{-1}$ for fixed $R$ and $\rLMCsidesig$ km  s$^{-1}$ for random $R$.
As canonical value we adopt the value $\sigma_\bd=4$ km s$^{-1}$,  known from stars in globular clusters.
The value $c_\hcc=1$ is also desired to avoid a too small $\sigma_\bd\approx4 \, c_\hcc^{1/2}$ km s$^{-1}$.

\subsection{On intensities}

At the time of this research, the intensities of the SMC map were not made public, so they were not considered.
By now the rough data are available on the Herschel site and clean data are to be produced from them (M. Meixner, private discussion).
In a future study we intend to incorporate that information for a further test of our model.

Here we look at a few characteristics. The Planck function gives the amount of electromagnetic energy radiated by a black body in
 thermal equilibrium per unit of time, frequency, area and solid angle,  $B_\nu(T)=2h\nu^3c^{-2}[{\exp(h\nu/k_BT)-1}]^{-1}$.
If the \mBD \ surface is optically thick, one expects a nearly exact Planck spectrum. The \mBDs \ in a given JC may have a distribution $\tau(T)$ of temperatures, 
which leads to $\bar B_\nu=\int\d T\tau(T)B_\nu(T)$.  In the modeling as cosmic dust, one often takes $\bar B_\nu\approx(\nu/\nu_0)^\beta B_\nu(T_0)$
with spectral index $\beta\approx2$ \cite{Juvela2010}. Here we continue with assuming a pure Planck spectrum.

The radiation into the outward hemisphere picks up an angular factor $\pi$.
The energy emitted per second per Hz by $N_\bd$ thermal spheres of radius $R_\bd$ is $4\pi^2 N_\bd B_\nu R_\bd^2$.
Measured per  unit area at distance $d$, this amounts to a specific intensity

\BEQ \label{J=Bnu}
\hspace{-2cm}
I_\nu=\pi B_\nu \frac{N_\bd R_\bd^2}{d^2}
=\pi B_\nu\frac{M_\jc R_\bd^2}{M_\bd d^2}=\pi B_\nu\frac{\sigma^2_\bd R_\jc R_\bd}{\sigma^2_\gas d^2}
=\pi B_\nu\frac{\sigma^2_\bd R_\bd}{\sigma^2_\gas R_\jc} \frac{\theta^2}{\textrm{sr}}.
\EEQ
In the last identities we model, next to the JCs, 
also the \mBDs \ as isothermal spheres~\cite{NSG2011}. 
They are supposed to consist of gas, H with the primordial He fraction $\phi=25\%$ in weight. Neglecting metals, this
 leaves $\rho_\H $ $=n_\H m_N=(1-\phi)\rho$.  With $p=(n_\H+n_\He)k_BT$, we denote $\sigma_\gas^2\equiv p/\rho=(1-\frac{3}{4}\phi)k_BT/m_N$, 
 which yields $\sigma_\gas=317\, \m\,\s^{-1}(T/15\K)^{1/2}$.

In a statistical model we may define the intrinsic JC intensity $i_\nu\equiv I_\nu d^2R^{-2}=I_\nu\theta^{-2}$, and assume local fluctuations with 
density $f_\nu(x,i_\nu)$ normalized as $\int\d i\, f_\nu(x,i)=f(x)$. 
This yields $\ddrie N/\d\theta\d\Omega\d I_\nu=A\theta^{-4}C_\nu(\theta,I_\nu\theta^{-2})$ with $C_\nu(\theta,i)$ from (\ref{CSun=}) with $f(x)\to f_\nu(x,i)$.

In the raw $250\mu$ Herschel data each pixel has size $\theta_\pix=6''$ and a saturation value between 0 and a few, in units of Jy beam$^{-1}$. 
We denote by $s_\jc$ the average saturation value of the pixels that expose the considered JC.
The effective beam area  is $\Omega_{\rm beam}=9.28 \,10^{-9}$ sr \cite{Meixner2011},
so the intensity recorded by a pixel with saturation $s_\jc $ is $91.2\,s_\jc $ mJy.
A JC of angular radius $\theta$ involves a number of pixels  $N_\pix=\pi\theta^2\theta_\pix^{-2}$ and an intensity 
$I_\nu= s_\jc \pi \theta^2\Omega_{\rm beam}^{-1}$ Jy.
Equating this with (\ref{J=Bnu}) yields an estimate for the \mBD \  radius,

\myskip{
\BEQ \hspace{-2cm}
R_\bd=\frac{0.286 s_\jc }{\pi\theta_\pix^2}\frac{\sigma_\gas^2}{\sigma_\bd^2}\frac{{\rm Jy}}{B_\nu}R_\jc
=70.1R_\odot\,\,\frac{s_\jc}{0.04} \, \frac{T}{15\K}\frac{B_\nu(15{\rm K})}{B_\nu(T)}\left( \frac{3.5\,\km\,\s^{-1}}{\sigma_\bd}\right)^2\frac{R_\jc}{2.5\,{\pc}},
\EEQ
\BEQ \hspace{-2cm}
R_\bd=\frac{91.2 s_\jc }{\theta_\pix^2}\frac{\sigma_\gas^2}{\sigma_\bd^2}\frac{{\rm mJy}}{B_\nu}R_\jc
=70.1R_\odot\,\,\frac{s_\jc}{0.04} \, \frac{T}{15\K}\frac{B_\nu(15{\rm K})}{B_\nu(T)}\left( \frac{3.5\,\km\,\s^{-1}}{\sigma_\bd}\right)^2\frac{R_\jc}{2.5\,{\pc}},
\EEQ}

\BEQ \hspace{-2cm}
R_\bd=\frac{s_\jc\textrm{ sr}}{\Omega_{\rm beam}}\,\frac{\textrm{Jy}}{B_\nu}\,\frac{\sigma_\gas^2}{\sigma_\bd^2}R_\jc
=53.6R_\odot\,\,\frac{s_\jc}{0.05} \, \frac{T}{15\K}\frac{B_\nu(15{\rm K})}{B_\nu(T)}\left( \frac{4\,\km\,\s^{-1}}{\sigma_\bd}\right)^2\frac{R_\jc}{2\,{\pc}},
\EEQ
i.e., dozens of solar radii. The \mBD \  mass then follows as a good dozen of Earth masses, 

\BEQ \label{Mbd=}
M_\bd=18.8\,M_\oplus\,\,\frac{s_\jc}{0.05} \, \left(\frac{T}{15\K}\right)^2\frac{B_\nu(15{\rm K})}{B_\nu(T)}\left( \frac{4\,\km\,\s^{-1}}{\sigma_\bd}\right)^2\frac{R_\jc}{2\,{\pc}}.
\EEQ
With typically $s_\jc\sim0.05$ this compares with the estimate of $13M_\oplus$ \cite{NGSEPL2009}, and is somewhat larger than the 
$\sim3 M_\oplus$ deduced  from quasar microlensing  observation \cite{Schild1996}.

We can also consider the microlensing of an SMC star by an \mBD \ inside  one of the JCs in front of the SMC, 
at distance $d=xd_{\rm SMC}$ with $x<1$. The Einstein radius reads:

\BEQ
\hspace{-2.25cm}
 R_E=\frac{2}{c}\sqrt{{x(1-x)GM_\bd d_{\rm SMC}}}=
 0.34R_\bd \,\frac{\sigma_\bd}{4\,\km\,\s^{-1}}\sqrt{\frac{4x(1-x)}{20s_\jc }\frac{B_\nu(T)}{B_\nu(15{\rm K})}\frac{2\,{\rm pc}}{R_\jc}}.
\EEQ
(For a star towards the Milky Way the factor 0.34 would be at most 0.12.)
Unless $T$ is well above 15 K,
one typically has $R_E<R_\bd$.  This is the case of {\it occultation}, where the big
\mBD \ absorbs the direct light when it passes in front of an SMC star, 
while its outer atmosphere refracts light \cite{SchildDekker}.
 With the \mBD \ speed dominated by that of the respective JC 
 in which it is embedded, the lensing-occultation event would last   $t_{\rm occ}\sim 2R_\bd/\vsun\sim3.5$ days.

\section{Conclusion}

Early in 2012, an intriguing multifrequency map of the Small Magellanic Cloud was made public. In its color presentation, it contains thousands of red dots,
Herschel cold clouds  (HCCs), related to the 250 micron observations that we present in Fig. 1. 

As a second year project at the University of Amsterdam, it was proposed to analyze the distribution of these HCCs using the  program ``ImageJ''.
With only the angular shapes available, no statement can be made about individual objects, so 
the approach has to embody a statistical analysis (discussed in section 4) and a comparison with theory.
Preliminary results were reported at the end of the term \cite{2ndyear1}.

In  the theoretical modeling we assume that the Galactic halo is essentially spherically symmetric, taking into account that the dark matter, whatever its nature,  
is probably well mixed.   A statistical analysis is promising, because it exhibits a scaling behavior which, if valid in some region, may apply to the full Halo.

Since we count many (6600) HCCs in the 50\% usable part of the  $\sim8^\circ\times 4^\circ$ SMC field, 
we assume that they represent a certain fraction of the total mass, and are well mixed too, 
so that their distribution has the same shape as the dynamical mass distribution. The Galaxy's slowly decaying circular rotation speed is modeled
as $v_c(r)=\vsun  (d_\SA/r)^\eps$  with $\eps=0.11$, which corresponds to the nearly isothermal ($\sim r^{-2-2\eps}$) mass density for the Galaxy, and, 
in the model, to an $r^{-2-2\eps}$ number density for the HCCs.
We allow that the density has also local fluctuations in the HCC size.
The number of HCCs per unit angular radius per  unit solid angle, $\dtwee N(\theta)/\d\theta\d\Omega$, 
behaves roughly as $\bA\theta^{-2}$ for $5''<\theta<30''$.
The amplitude $\bA\sim100$ is reasonable,  taking into account that typically $\theta\sim20''$.
This corresponds to an amount of HCCs in the field of Fig. 1 of the order 
$8^\circ\times4^\circ\times \d N/\d\Omega\sim 0.01 \theta\, \dtwee N/\d \theta\d\Omega\sim 0.01 A/\theta\sim10,000$,
indeed about twice the number 6600 of HCCs with $N_\pix\ge6$  that we count up to 40 kpc in the usable areas.

Another intuitive interpretation of our numbers is as follows.
If we take our SMC count of 6600 HCCs, multiply this by 5 for covering the Halo up to 200 kpc
and then by $4 \pi/(50\%\times 8.3^\circ\times 4.1^\circ)=2400$ for covering the whole sky, 
we arrive at an expected total number of $7.9\,10^7$ HCCs for the Galaxy;
if they weigh $10,000 M_\odot$ on average, this constitutes $0.79\, 10^{12}M_\odot$ in total.
This rough estimate stems  with the expected total Galactic mass, 
and can be compared with  the $1.1\,10^{12}M_\odot$ from Eq. (2), that is incorporated
in our detailed modeling of the spatial HCC distribution.

Seen from the Earth,  nearby objects look larger than from \SagA, which induces a depletion of the HCC histogram towards larger angular radii, 
up to a factor 2 at $30''$. By itself this has little statistical meaning, since our error bars are large. 
However, our distance to \SagA \ is big enough to yield a sizable effect in the prediction as well as a good  fit for the average HCC radius. 
The  case of equal radii brings $\bar R=1.7$ pc; for comparison,  the average half-light radius of 157 GCs
 in the Milky Way is 4.3 pc \cite{HarrisGC}; being mostly heavier, GCs probably arise from agglomerates of HCCs.
This value of $\bar R$ sets the distance scale of Figs. 1 and 2, and relates the investigated angles to distances between 10 and 40 kpc.
The HCCs bigger than the diffraction limit appear to be located inside the Galactic Halo, in front of the SMC.
The less informative LMC map provides 
results roughly consistent with those for the SMC.

 Our SMC-LMC analysis is compatible with a nearly isothermal distribution of Herschel cold clouds.
 If they are distributed in a spherically symmetric fashion throughout the Halo, they can explain the full dynamical mass of the Galaxy, 
 provided their average mass is of the order of 10,000 $M_\odot$.
 One is thus led to identify the HCCs of heretofore unknown origin with the Jeans clusters (JCs) predicted by the Jeans instability. 
Our mass value is in accord with the Jeans instability, which is known to have produced globular star clusters, probably from mergers of JCs.
The value of 10,000 $M_\odot$ is even somewhat smaller than might be expected, which may be due to an over-concentration of Jeans clusters in front 
 of the Magellanic clouds. Indeed, some clustering, that we observe at small scale,  should also be expected on larger scale.

Thousands of similar cold clouds have been detected by Planck \cite{PlanckVII} and by Herschel/Spitzer \cite{Wilcock2012}.
Both the parsec size and 10,000 $M_\odot$ weight of the supposed Jeans clusters are in accord with the
size and gas masses of some of these clouds \cite{Juvela2010,Juvela2012,Wilcock2012}.

The amplitude $A$ fixes the ratio $\bar M_\jc/\bar R_\jc$, and since in Gravitational Hydrodynamics the JCs themselves
are isothermal spheres of micro brown dwarfs (\mBDs) of Earth weight scale,  it determines
 their velocity dispersion as $\sigma_\bd\approx4$ km s$^{-1}$,
a value known from stars in globular star clusters, in support of their presumed connection to Jeans clusters.

Another surprise is that the deduced $\sigma_\bd$  is close to the speed of sound at recombination $5.7$ km s$^{-1}$,
the early moment in the history of the Universe (at  377.000 years after the Big Bang) when the plasma converted into gas. 
Intriguingly, the \mBDs \ still seem to move at nearly the speed of sound of the gas that fragmented to create them!

Our analysis points at occultation rather than lensing, which may explain why \mBDs \ (MACHOs) can exist but were not observed 
in microlensing towards the LMC, SMC and the Milky Way \cite{Renault1998,Alcock2000}:
 {\it they may not act as lenses but as refractive dark objects}.  An occultation  event in Centaurus A has lasted 170 days \cite{Rivers2011}.
The cloud's mass of 3 - 10 $M_\oplus$ agrees with (\ref{Mbd=});  its size of Pluto-orbit scale 
can be caused by heating of the \mBD~\cite{Gibson1996}.
A documented doubly-peaked quasar microlensing event \cite{ColleySchild2003}  can be explained as lensing with partial occultation, 
due to a cloud of about an Earth mass and a size of a few solar radii, located in the Halo. This  support our present findings,  including occultation 
towards the MCs \cite{NColleySchild2012}.
 
We conclude that, under the assumption of a spherically symmetric Halo,
{\it enough cold clouds have been observed towards the Magellanic clouds for the Jeans mechanism to explain the 
full dynamical mass of the Galaxy}.
In other words, {\it cold JCs can explain the dark matter and most of the missing baryons of the Galaxy}.
The obvious implication is that CDM would not be needed to explain dynamical  behavior in the Galaxy and, by default, at galactic scales.
This conclusion was reached before on a the basis of another field of the sky and by other arguments~\cite{NSG2011}. 
If CDM does not occur in the Galaxy, then it likely does not exist at all. But this raises the question: What is the true non-baryonic dark matter?
A fit to lensing data of the galaxy cluster Abell 1689 has shown that 1.5 eV neutrinos answer the question \cite{NEPL2009}.
This mass exceeds estimates based on the cold dark matter paradigm, but lies below the upper bound of 2 eV.
It will be tested in the upcoming KATRIN tritium-decay experiment~\cite{Wong2011}.

As an outlook, the full data for the SMC will also contain the intensities of the HCCs. 
Investigation of other fields on the sky seems feasible. A dedicated MACHO search towards the LMC can look for refractive 
lensing-occultation events.

\subsection*{Acknowledgements}
We thank for discussion with Rudy E. Schild, Katerina Dohnalov\'a and Ian Avruch.

\newcommand{\MNRAS}{Monthly Notices of the Royal Astronomical Society}

\subsection*{References}

\newcommand{\apj}{Astrophys. J.}
\newcommand{\aj}{Astron. J.}
\newcommand{\asas}{Astron. \& Astroph.}

\newpage

\end{document}